\documentclass[letterpaper,twocolumn,american,showpacs,pra,aps,superscriptaddress]{revtex4}
\usepackage[latin1]{inputenc}
\usepackage{bm}
\usepackage{multirow,amssymb,amsbsy,amsmath}
\usepackage{stmaryrd}
\usepackage{graphicx}
\usepackage{hyperref}
\makeatletter
\usepackage{pifont}
\makeatother

\begin{document}

\title{Correlated projection operator approach to non-Markovian dynamics in spin baths}

\author{Jan Fischer}

\email{jan.fischer@unibas.ch}

\affiliation{Physikalisches Institut, Universit\"at Freiburg,
             Hermann-Herder-Strasse 3, D-79104 Freiburg, Germany}
\affiliation{Department of Physics, University of Basel, 
			Klingelbergstrasse 82, CH-4056 Basel, Switzerland}

\author{Heinz-Peter Breuer}

\email{breuer@physik.uni-freiburg.de}

\affiliation{Physikalisches Institut, Universit\"at Freiburg,
             Hermann-Herder-Strasse 3, D-79104 Freiburg, Germany}

\date{\today}

\begin{abstract}
The dynamics of an open quantum system is usually studied by
performing a weak-coupling and weak-correlation expansion in the
system-bath interaction. For systems exhibiting strong couplings
and highly non-Markovian behavior this approach is not justified.
We apply a recently proposed correlated projection superoperator
technique to the model of a central spin coupled to a spin bath
via full Heisenberg interaction. Analytical solutions to both the
Nakajima-Zwanzig and the time-convolutionless master equation are
determined and compared with the results of the exact solution.
The correlated projection operator technique significantly
improves the standard methods and can be applied to many physical
problems such as the hyperfine interaction in a quantum dot.
\end{abstract}

\pacs{03.65.Yz, 42.50.Lc, 03.65.Ta, 73.21.La}

\maketitle

\section{Introduction}

Some of the key features in the dynamics of open quantum systems
\cite{OpenQS} are phenomena such as relaxation, decoherence and
the buildup of correlations and entanglement due to the
interaction of the open system with its environment. This behavior
is generic in the sense that it is observable for many different
kinds of environments and microscopic interactions. For many
physical systems it is justified to assume that the coupling to
the environment is weak (Born approximation) or that correlations
in the bath decay quickly with respect to the typical timescale of
the system's dynamics (Markov approximation). In this case it is
often possible to construct a generator for the dynamics which is
in Lindblad form \cite{LINDBLAD,GORINI}.

In general, however, memory effects in the bath cannot be
neglected and the Markov assumption is not applicable any more.
Such systems are said to exhibit non-Markovian behavior. This can
be due to strong system-environment couplings \cite{REIBOLD,HU},
correlations and entanglement in the initial state
\cite{INGOLD,BUZEK}, finite reservoirs \cite{GEMMER2005,GMM}, or
due to coupling to environments at low temperatures or to spin
baths \cite{LOSS,COISH,STOLZE}. Also, heat transport in
nano-structures \cite{FOURIER} has been shown to exhibit
non-Markovian behavior.

A powerful tool for dealing with such systems is provided by the
projection operator techniques \cite{HAAKE, KUBO} which have been
introduced by Nakajima \cite{NAKAJIMA}, Zwanzig \cite{ZWANZIG} and
Mori \cite{MORI}. These techniques are based on the introduction
of a projection superoperator $\mathcal{P}$ which acts on the
density operator $\rho$ of the total system and projects onto the
so-called relevant part $\mathcal{P} \rho$. The latter represents
a certain approximation of $\rho$, and this procedure leads to a
simplified effective description of the dynamics through a reduced
set of relevant variables. Stated differently, the projection
$\mathcal{P}$ expresses the elimination of certain degrees of
freedom which are considered negligible for the treatment of the
open system's dynamics.

Here, we will consider two different approaches which both lead to a closed
equation for the dynamics of the relevant part of $\rho$. The first one
leads to the Nakajima-Zwanzig equation \cite{NAKAJIMA, ZWANZIG}, an
integrodifferential master equation for $\mathcal{P} \rho$ containing a
memory kernel. The second approach eliminates the integration over the
system's history and leads to the time-convolutionless master equation
which is local in time, but involves a time-dependent generator
\cite{SHIBATA77, SHIBATA79, SHIBATA80, SHIBATA99, ROYER2003, BREUER2004}.
In both cases the resulting master equation is used as a starting point for a
systematic expansion in powers of the system-environment interaction.

Usually the projector is chosen such that $\mathcal{P} \rho =
(\mathrm{tr}_E \rho) \otimes \rho_0 = \rho_S \otimes \rho_0$,
where $\rho_0$ is some fixed environmental state, e.g.~a thermal
equilibrium state, and where $\mathrm{tr}_E$ denotes the partial
trace over the environment. The superoperator $\mathcal{P}$
projects the total state $\rho$ onto a tensor-product state,
i.e., onto a state without any statistical correlations between
the open system and its environment. If correlations remain small,
this ansatz is justified and the perturbation expansion is usually
carried out only to leading order in the perturbation (Born
approximation). However, for many physical systems exhibiting
strong correlations, the Born approximation might be inadequate
such that one also has to take into account higher orders of the
expansion. The calculation of higher orders is limited by the
increase of mathematical complexity, though, and furthermore, the
expansion may not converge uniformly in time, such that higher
orders may diverge on longer timescales \cite{BURGARTH}.

A different strategy for taking strong correlations into account
is to introduce a superoperator $\mathcal{P}$ that projects onto a
correlated system-environment state, i.e., onto a state that
contains certain statistical correlations between the open system
and its environment
\cite{CPS,BGM,BUDINI06,BUDINI05,SCHOMERUS,GASPARD1,GASPARD2}. We
will refer to such $\mathcal{P}$ as a correlated projection
superoperator. In the present article we will discuss in detail
the application of the correlated projection operator method to a
two level system interacting with a spin bath. This system is
particularly interesting since it models the hyperfine interaction
of an electron confined to a quantum dot with its surrounding
nuclei \cite{COISH}.

We organize this paper as follows. In
Sec.~\ref{STANDARD-PROJECTION-OPERATORS} we review the standard
projection operator technique and formulate the Nakajima-Zwanzig
as well as the time-convolutionless master equation. Also, we
briefly recapitulate how to extend the technique to correlated
projection superoperators. In Sec.~\ref{APPLICATIONS} we apply
this method to a simple spin-bath model and construct projectors
on the basis of symmetry considerations that vastly improve the
performance of the projection operator technique. We compare the
resulting solutions of the Nakajima-Zwanzig and the
time-convolutionless master equation in second order to an exact
solution of the model. Discussion and conclusions are then
provided in Sec.~\ref{CONCLU}.

\section{Projection operator techniques}\label{STANDARD-PROJECTION-OPERATORS}

In general, one considers an open system $S$ which is coupled to
some environment $E$. The accordant Hilbert spaces are denotes by
$\mathcal{H}_S$ and $\mathcal{H}_E$, respectively, and the state
space of the total system is given by the product $\mathcal{H} =
\mathcal{H}_S \otimes \mathcal{H}_E$. The Hamiltonian of the total
system is denoted by $H = H_0 + H_I$, where $H_0$ is the
unperturbed part of the Hamiltonian and $H_I$ represents the
interaction between system and bath. A general state of the total
system is given by a density matrix $\rho$. The partial traces
over the system $S$ and the environment $E$ are denoted by
$\mathrm{tr}_S$ and $\mathrm{tr}_E$, respectively. The reduced
density matrix of the open system is thus given by
$\rho_S=\mathrm{tr}_E\rho$.

\subsection{The standard projection superoperator}

The starting point of the projection operator techniques is the introduction
of a superoperator $\mathcal{P}$ which acts on the total system's density
matrix, and which is defined by
\begin{equation} \label{STANDARD-PROJECTION}
    \mathcal{P} \rho = (\mathrm{tr}_E\rho) \otimes \rho_0,
\end{equation}
where $\rho_0$ is some fixed state of the environment. The map
$\mathcal{P}$ satisfies the condition of a projector, namely
${\mathcal{P}}^2={\mathcal{P}}$, and is therefore referred to as a
projection superoperator, being a map acting on operators. The
complementary map is defined via $\mathcal{Q} = I - \mathcal{P}$,
where $I$ denotes the identity. Note that $\mathcal{P} \rho$
contains all information about the open system in the sense that
for the expectation value of any observable $\mathcal{O}_S$ of the
open system the relation $\mathrm{tr} \{ \mathcal{O}_S \rho \} =
\mathrm{tr} \{ \mathcal{O}_S \mathcal{P} \rho \}$ holds.

\subsection{Nakajima-Zwanzig equation}

Starting from the von Neumann equation in the interaction picture with
respect to $H_0$,
\begin{equation} \label{NEUMANN-EQ}
 \frac{d}{d t} \rho(t) = -i [H_I(t),\rho(t)] \equiv \mathcal{L}(t) \rho(t),
\end{equation}
one can derive a closed equation for the projection
${\mathcal{P}}\rho(t)$ by inserting $\mathcal{P}$ and
$\mathcal{Q}$ in front of $\rho$ on both sides,
\begin{equation} \label{NZ-GEN}
    \frac{d}{dt} \mathcal{P} \rho(t) = \int_0^t dt_1 \;
    \mathcal{K}(t,t_1) \mathcal{P} \rho(t_1).
\end{equation}
The superoperator
\begin{equation} \label{NZ-KERNEL}
    \mathcal{K}(t,t_1) = \mathcal{P} \mathcal{L}(t) \>
    \mathrm{T} \exp \left[ \int_{t_1}^t dt_2 \mathcal{Q} \mathcal{L}(t_2) \right]
    \mathcal{Q} \mathcal{L}(t_1) \mathcal {P}
\end{equation}
is referred to as the memory kernel or the self-energy, and the
operator $\mathrm{T}$ denotes chronological time-ordering.
Eq.~(\ref{NZ-GEN}) is called the Nakajima-Zwanzig (NZ) equation
and describes non-Markovian behavior of the total system through
the memory kernel (\ref{NZ-KERNEL}). In general, there is an
additional term proportional to $\mathcal{Q} \rho(0)$ on the
right-hand side of the NZ equation which we have omitted here for
simplicity, supposing an initial state that satisfies $\mathcal{P}
\rho(0) = \rho(0)$. Since $\mathcal{K}(t,t_1)$ is usually a very
complicated operator, it is customary to perform a perturbation
expansion in powers of $H_I$. Under the condition
\begin{equation}\label{PLP-ASSUM}
 {\mathcal{P}}{\mathcal{L}}(t){\mathcal{P}} = 0
\end{equation}
the lowest-order contribution is given by the second order,
\begin{equation} \label{NZ2}
    \frac{d}{d t} {\mathcal{P}}\rho(t) = \int_0^t d t_1
    {\mathcal{P}}{\mathcal{L}}(t){\mathcal{L}}(t_1){\mathcal{P}}\rho(t_1),
\end{equation}
and higher orders are obtained by expanding the time-ordered
exponential of the memory kernel (\ref{NZ-KERNEL}).

\subsection{Time-convolutionless master equation}\label{TCL-TECHNIQUE}

An alternative way of deriving an exact master equation for the
relevant part of $\rho$ is to remove the dependence of the
system's dynamics on the full history of the system and to
formulate a time-local equation of motion, which is given by
\begin{equation} \label{TCL-GEN}
    \frac{d}{d t} {\mathcal{P}}\rho(t) = {\mathcal{K}}(t)
    {\mathcal{P}}\rho(t).
\end{equation}
This equation is called the time-convolutionless (TCL) master
equation, and $\mathcal{K}(t)$ is a time-dependent superoperator,
which is referred to as the TCL generator. As for the NZ equation,
in general there is also a term proportional to $\mathcal{Q}
\rho(0)$ on the right-hand side of Eq.~(\ref{TCL-GEN}), which
vanishes provided one chooses a factorized initial state.

Like for the NZ equation one can carry out a perturbation
expansion of the TCL generator in powers of $H_I$. The various
orders of this expansion can be expressed through the ordered
cumulants \cite{KUBO63,ROYER72,KAMPEN1,KAMPEN2} of the Liouville
superoperator $\mathcal{L}(t)$. The second-order contribution
reads
\begin{equation}
    {\mathcal{K}}_2(t) = \int_0^t d t_1
    {\mathcal{P}}{\mathcal{L}}(t){\mathcal{L}}(t_1){\mathcal{P}},
\end{equation}
such that the second-order TCL master equation takes the form
\begin{equation} \label{TCL2}
    \frac{d}{d t} {\mathcal{P}}\rho(t) = \int_0^t d t_1
    {\mathcal{P}}{\mathcal{L}}(t){\mathcal{L}}(t_1){\mathcal{P}}\rho(t),
\end{equation}
which should be contrasted to the second-order NZ equation
(\ref{NZ2}). One can formulate a simple set of rules that enable
one to write down immediately these expressions and the corresponding expressions
for all higher orders \cite{OpenQS}. The TCL technique has been
applied to many physical systems. Examples, which include the
determination of higher orders of the expansion, are the damped
Jaynes-Cumings model \cite{BKP}, quantum Brownian motion and the
spin-boson model \cite{DECOHERENCE}, the spin-star model
\cite{BURGARTH}, relaxation processes in structured reservoirs
\cite{BGM}, and the dynamics of the atom laser \cite{ATOMLASER}.
Applications of the TCL method to systems relevant for quantum
information processing may be found in \cite{AHN02,AHN05}.
Recently, the performance of the TCL approach has been studied for
the model of a qubit coupled to a spin bath via Ising interaction
\cite{LIDAR07}, and for a general open system coupled to a bath
through pure phase noise \cite{HAENGGI}.

We emphasize at this point that the TCL master equation describes
non-Markovian dynamics although it is local in time. All memory
effects are taken into account by the explicit time-dependence of
the TCL generator. It should also be pointed out that both the NZ
equation (\ref{NZ-GEN}) and the TCL equation (\ref{TCL-GEN}) are
exact in the sense that they are equivalent to the equation of
motion gained from the full system Hamiltonian.

It is important to realize that the NZ and the TCL techniques lead
to equations of motion with entirely different structures.
Therefore, also the mathematical structure of their solutions are
quite different in any given order \cite{ROYER03}. It turns out
that in many cases the degree of accuracy obtained by both methods
is of the same order of magnitude. In these cases the TCL approach
is of course to be preferred because it is technically much
simpler to deal with.

\subsection{Correlated projection superoperators}\label{SEC:CORR}

The performance of the projection operator techniques depends on
the properties of the microscopic model under study, in particular
on the structure of the environmental correlation functions.
However, it also depends strongly on the choice of the
superoperator ${\mathcal{P}}$. Several extensions of the standard
projection (\ref{STANDARD-PROJECTION}) and modifications of the
expansion technique have been proposed in the literature (see,
e.g., \cite{OPPENHEIM1,OPPENHEIM2,FRIGERIO}).

A characteristic feature of the projection defined by
Eq.~(\ref{STANDARD-PROJECTION}) is given by the fact that it
projects any state $\rho$ onto a tensor product $\rho_S \otimes
\rho_0$ that describes a state without statistical correlations
between the system and its environment. This is not the only
possible choice. In fact, a general class of projection
superoperators can be represented as follows,
\begin{equation} \label{PROJECTION-GENFORM}
 {\mathcal{P}}\rho = \sum_i {\mathrm{tr}}_E \{ A_i \rho \}
 \otimes B_i,
\end{equation}
where $\{A_i\}$ and $\{B_i\}$ are two sets of linear independent
Hermitian operators on ${\mathcal{H}}_E$ satisfying the relations
\begin{eqnarray}
 {\mathrm{tr}}_E \{ B_i A_j \} &=& \delta_{ij}, \label{BjAi} \\
 \sum_i ({\mathrm{tr}}_E B_i) A_i &=& I_E, \label{TRACE-PRESERVING} \\
 \sum_i A_i^T \otimes B_i &\geq& 0. \label{COND-POS}
\end{eqnarray}
The map defined by Eq.~(\ref{PROJECTION-GENFORM}) projects in
general onto correlated (non-factorizing) system-environment
states and may thus be referred to as a correlated projection
superoperator. It can be shown \cite{CPS} that
Eq.~(\ref{PROJECTION-GENFORM}) represents the most general form of
such a correlated projection superoperator under certain natural
physical conditions. These conditions demand that ${\mathcal{P}}$
is a completely positive and trace preserving map (quantum
channel) that operates on the environmental variables. Equation
(\ref{BjAi}) guarantees that ${\mathcal{P}}$ is a projection
superoperator, Eq.~(\ref{TRACE-PRESERVING}) ensures that
${\mathcal{P}}$ preserves the trace (normalization), while
Eq.~(\ref{COND-POS}) is equivalent to the condition of complete
positivity ($T$ denotes the transposition). We remark that the
class of projections defined by Eq.~(\ref{PROJECTION-GENFORM})
yields a natural generalization of the Lindblad equation to the
regime of strong non-Markovian quantum dynamics
\cite{CPS,BUDINI06}.

A typical example for a correlated projection that is relevant for
the present paper is obtained by the choice
\begin{equation}
 A_i = \Pi_i, \qquad B_i = \frac{1}{N_i}\Pi_i,
 \qquad N_i = {\mathrm{tr}}_E \Pi_i,
\end{equation}
where the $\Pi_i$ are ordinary projection operators on
${\mathcal{H}}_E$ which form an orthogonal decomposition of the
unit operator:
\begin{equation}
 \Pi_i \Pi_j = \delta_{ij}\Pi_j, \qquad \Pi_i^{\dagger} = \Pi_i,
 \qquad \sum_i \Pi_i = I_E.
\end{equation}
One easily checks that with this choice all conditions formulated
above are indeed satisfied. Further examples are discussed in
Refs.~\cite{BGM,CPS,BUDINI06,GASPARD1,GASPARD2}.

As mentioned already in Sec.~\ref{TCL-TECHNIQUE}, a homogeneous NZ
or TCL equation of motion presupposes a tensor product initial
state if one uses the standard projection superoperator
(\ref{STANDARD-PROJECTION}). However, this is no longer true for
the correlated projection defined by
Eq.~(\ref{PROJECTION-GENFORM}). In fact, the general condition for
the absence of an inhomogeneous term in the NZ equation
(\ref{NZ-GEN}) or the TCL equation (\ref{TCL-GEN}) is given by
${\mathcal{P}}\rho(0)=\rho(0)$. According to
Eq.~(\ref{PROJECTION-GENFORM}) this condition does {\textit{not}}
require that $\rho(0)$ be a factorized state. Hence, a great
advantage of the correlated projection superoperators is given by
the fact that they allow the treatment of correlated initial
states by means of a homogeneous NZ or TCL equation
\cite{CPS,Durban-Proc}.

Once $\mathcal{P}$ is chosen, the dynamics of the open system is
uniquely determined by the dynamical variables
\begin{equation}
    \rho_i(t) = \mathrm{tr}_E \{ A_i \rho(t) \}.
\end{equation}
The connection to the reduced density is simply given by
\begin{equation}\label{rhosum}
    \rho_S(t) = \sum_i \rho_i(t),
\end{equation}
and the normalization condition reads
\begin{equation}
    \mathrm{tr}_S \> \rho_S(t) = \sum_i \mathrm{tr}_S \> \rho_i(t) = 1.
\end{equation}
The reduced density matrix is hence uniquely determined by a set
of (unnormalized) operators $\rho_i(t)$. With the help of the NZ
equation (\ref{NZ-GEN}) one finds that the dynamics of these
operators is governed by a coupled system of integrodifferential
equations,
\begin{equation} \label{NZ-RHO-I}
 \frac{d}{dt} \rho_i(t) = \sum_j \int_0^t dt_1 \;
 \mathcal{K}_{ij}(t,t_1) \rho_j(t_1),
\end{equation}
where the superoperators $\mathcal{K}_{ij}(t,t_1)$ are defined
through their action on an arbitrary system operator
${\mathcal{O}}_S$,
\begin{equation}
 \mathcal{K}_{ij}(t,t_1) \mathcal{O}_S \equiv
 {\mathrm{tr}}_E \left\{ A_i \mathcal{K}(t,t_1) (\mathcal{O}_S \otimes B_j) \right\}.
\end{equation}
On the other hand, the TCL equation (\ref{TCL-GEN}) leads to a
coupled system of time-local differential equations,
\begin{equation} \label{TCL-RHO-I}
 \frac{d}{dt} \rho_i(t) = \sum_j \mathcal{K}_{ij}(t) \rho_j(t),
\end{equation}
with superoperators defined by
\begin{equation}
 \mathcal{K}_{ij}(t) \mathcal{O}_S \equiv
 {\mathrm{tr}}_E \left\{ A_i \mathcal{K}(t) (\mathcal{O}_S \otimes B_j) \right\}.
\end{equation}

A crucial step for a successful applications of the correlated
projection operator technique is the construction of an
appropriate projection superoperator ${\mathcal{P}}$. A useful
strategy for this construction is to take into account the known
conserved quantities of the model under study. Let us suppose that
$C$ is a such conserved quantity. Then a good choice for
$\mathcal{P}$ is a projection that leaves invariant the
expectation values of $C$, i.e., $\mathrm{tr}\{C\rho\}=
\mathrm{tr}\{C\mathcal{P}\rho\}$. Requiring this to hold for all
states $\rho$ we get the condition
\begin{equation} \label{C-COND}
 {\mathcal{P}}^{\dagger} C = C,
\end{equation}
where ${\mathcal{P}}^{\dagger}$ denotes the adjoint of
${\mathcal{P}}$ with respect to the Hilbert-Schmidt scalar product
\cite{Durban-Proc}. This equation represents a condition for the
projection superoperator ${\mathcal{P}}$ on the basis of a known
conserved quantity of the underlying model. It ensures that the
projection superoperator leaves invariant this quantity and that
the effective description respects the corresponding conservation
law \cite{GASPARD1,GASPARD2}.

\section{Application to a spin bath}\label{APPLICATIONS}

\subsection{Description of the model}\label{SPIN-STAR-MODEL}

We want to apply the correlated projection operator formalism to a
spin-bath model, which is defined by the Hamiltonian
\begin{equation} \label{H-STAR}
    H = \frac{\omega_0}{2} \sigma_3 +
    \sum_{k=1}^{N} A_k \bm{\sigma}\cdot\bm{\sigma}^{(k)}.
\end{equation}
The central spin, representing the open system, has an energy
splitting $\omega_0$ and is coupled to the $k$-th bath spin via
the coupling constant $A_k$. Here, $\bm{\sigma}$ and
$\bm{\sigma}^{(k)}$ are the Pauli operators of the central spin
and of the $k$-th bath spin, respectively.

\begin{figure}[htb]
\begin{center}
\includegraphics[width=0.6\linewidth]{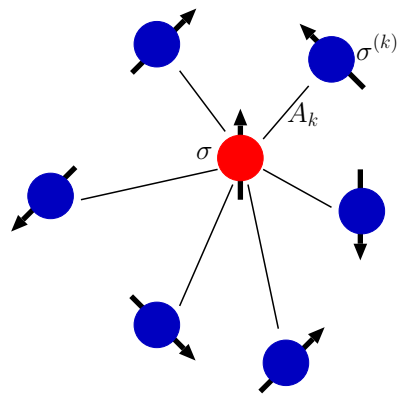}
\caption{(Color online) Illustration of the model
with Hamiltonian (\ref{H-STAR}). A central spin $\bm{\sigma}$
interacts with a bath of $N$ spins $\bm{\sigma}^{(k)}$ through the
coupling constants $A_k$.} \label{fig1}
\end{center}
\end{figure}

In the following we will restrict ourselves to initial states of
the form
\begin{equation} \label{INITIAL-STATE}
 \rho(0) = \rho_S(0) \otimes \frac{1}{2^N} I_E,
\end{equation}
where $I_E$ is the unit matrix in the state space of the spin
bath. Initially the bath is thus in a completely unpolarized
(infinite temperature) state. This choice is made here only for
simplicity; the present method also allows the treatment of
polarized bath states and of correlated initial states.

Our aim is to illustrate the TCL and the NZ expansion for a
correlated projection superoperator. To compare the results with
the exact solution of the full von Neumann equation corresponding
to the Hamiltonian (\ref{H-STAR}) we take the coupling constants
to be independent of $k$, i.e., $A_k = A$. The model can then be
solved analytically as follows. First one notes that the
3-component of the angular momentum of the total system,
\begin{equation}
 J_3^{\mathrm{tot}} = \frac{1}{2}\sigma_3 + J_3,
\end{equation}
is conserved under the time-evolution given by the Hamiltonian
(\ref{H-STAR}). Here, $J_3$ denotes the 3-component of the angular
momentum of the bath which is defined by
\begin{equation}
 \bm{J} = \frac{1}{2} \sum_k \bm{\sigma}^{(k)}.
\end{equation}
For $A_k=A$, also the square $\bm{J}^2$ of the bath angular
momentum is conserved. We introduce basis states $|j,m\rangle$
which are defined as simultaneous eigenstates of $\bm{J}^2$ and
$J_3$ \footnote{More precisely, these states should be written as
$|j,m,\lambda\rangle$, where $\lambda$ stands for additional
quantum numbers which, together with $j$ and $m$, uniquely fix the
basis state.}. If $N$ is even, $j$ takes on the values
$j=0,1,2,\ldots,\frac{N}{2}$, while
$j=\frac{1}{2},\frac{3}{2},\ldots,\frac{N}{2}$ for odd $N$. For a
given value of $j$, the quantum number $m$ takes on the values
$m=-j,-j+1,\ldots,+j$.

With the help of the conserved quantities one easily solves the
time-dependent Schr\"odinger equation, employing the fact that the
two-dimensional subspaces spanned by the states
$|+\rangle\otimes|j,m\rangle$ and $|-\rangle\otimes|j,m+1\rangle$
are invariant under the time-evolution. Here, the states
$|\pm\rangle$ denote the eigenstates of the central spin,
$\sigma_3|\pm\rangle=\pm|\pm\rangle$. We further introduce the
quantities
\begin{equation} \label{DEF-P-J}
 p(j) = \frac{2j+1}{2^N} N_j,
\end{equation}
where
\begin{equation} \label{Def-N-J}
 \qquad N_j =
 \binom{N}{\frac{N}{2}+j}-\binom{N}{\frac{N}{2}+j+1}
\end{equation}
represents the number of times the angular momentum $j$ appears in
the decomposition of the bath Hilbert space into irreducible
subspaces of the rotation group. Hence, $p(j)$ may be interpreted
as the probability of finding a certain angular momentum $j$ in
the initial state of the bath. Finally, we define the frequencies
\begin{eqnarray*}
 \Omega_{\pm}(m) &=& \pm\omega_0 + 4A(\pm m + 1/2), \\
 \mu_{\pm}(j,m) &=& \sqrt{\frac{\Omega^2_{\pm}(m)}{4}+4A^2[j(j+1)-m(m\pm 1)]}.
\end{eqnarray*}
With the help of these definitions the upper state population of
the central spin can be written as
\begin{eqnarray}
 P_+(t) &=& \langle +|\rho_S(t)|+\rangle \nonumber \\
 &=& \sum_{jm} \frac{p(j)}{2j+1} \Big[ \cos^2[\mu_+(j,m)t] \nonumber \\
 &~& \qquad + \frac{\Omega_+^2(m)}{4\mu_+^2(j,m)}
 \sin^2[\mu_+(j,m)t] \Big],
\end{eqnarray}
where we have used for simplicity the initial condition
$P_+(0)=1$. The lower state population is found from
$P_-(t)=\langle -|\rho_S(t)|-\rangle=1-P_+(t)$. The coherence of
the central spin is determined by the formula
\begin{eqnarray*}
 \rho_{+-}(t) &=& \langle +|\rho_S(t)|-\rangle \\
 &=& \rho_{+-}(0)
 \sum_{jm} \frac{p(j)e^{i \omega_0t}}{2j+1} \\
 &\times&
 \left[ \cos[\mu_+(j,m)t] - \frac{i \Omega_+(m)}{2\mu_+(j,m)} \sin[\mu_+(j,m)t] \right]
 \\ &\times&
 \left[ \cos[\mu_-(j,m)t] + \frac{i \Omega_-(m)}{2\mu_-(j,m)} \sin[\mu_-(j,m)t] \right].
\end{eqnarray*}

\subsection{Correlated projection operators}

According to Sec.~\ref{SEC:CORR}, a good candidate for a correlated
projection superoperator is a map that does not change the
expectation values of the known conserved quantities. We start by
using a projection superoperator ${\mathcal{P}}$ that leaves
invariant the 3-component of the total angular momentum, i.e.,
that satisfies [see Eq.~(\ref{C-COND})]
\begin{equation} \label{INVAR}
 {\mathcal{P}}^{\dagger}J^{\mathrm{tot}}_3=J^{\mathrm{tot}}_3,
\end{equation}
We define bath operators $\Pi_m$ which project onto the subspace
spanned by the eigenstates of $J_3$ corresponding to the
eigenvalue $m=-\frac{N}{2},\ldots,+\frac{N}{2}$. An appropriate
projection operator with the property (\ref{INVAR}) is then given
by
\begin{equation} \label{CPS1}
 {\mathcal{P}}\rho = \sum_m {\mathrm{tr}}_E \{ \Pi_m \rho\}
 \otimes \frac{1}{N_m} \Pi_m
 \equiv \sum_m \rho_m \otimes \frac{1}{N_m} \Pi_m,
\end{equation}
where the number
\begin{equation}
 N_m = {\mathrm{tr}}_E \Pi_m = \binom{N}{\frac{N}{2}+m}
\end{equation}
represents the degree of degeneracy of the eigenvalue $m$ of
$J_3$. One can easily check that this projection operator
satisfies all conditions formulated in Sec.~\ref{SEC:CORR}.

The state of the central spin is thus determined by the set of
densities $\rho_m(t)$. The projection ${\mathcal{P}}\rho$
represents a correlated system-environment state in which each
$\rho_m$ is correlated with the bath state $\Pi_m/N_m$ which
describes a state of maximal entropy under the constraint of a
given value $m$ for the 3-component of the angular momentum.
According to Eq.~(\ref{rhosum}) the reduced density matrix is
obtained from
\begin{equation} \label{RED-DENSITY}
 \rho_S(t) = \sum_m \rho_m(t),
\end{equation}
and its normalization condition reads
\begin{equation}
 {\mathrm{tr}}_S \; \rho_S(t) = \sum_m {\mathrm{tr}}_S \; \rho_m(t) = 1.
\end{equation}
Finally, the matrix elements of the $\rho_m(t)$ are defined by
\begin{eqnarray}
 P^m_{\pm}(t) &=& \langle \pm|\rho_m(t)|\pm\rangle, \\
 \rho^m_{+-}(t) &=& \langle +|\rho_m(t)|-\rangle.
\end{eqnarray}
It should be emphasized that the above procedure is also
applicable if the coupling constants $A_k$ are not equal to each
other, since $J_3^{\mathrm{tot}}$ is then still a conserved
quantity. We remark that the standard procedure with a projection
operator of the form of Eq.~(\ref{STANDARD-PROJECTION}) with
$\rho_0=2^{-N}I_E$ would give
${\mathcal{P}}^{\dagger}J^{\mathrm{tot}}_3=\frac{1}{2}\sigma_3$.

\subsubsection{Interaction Hamiltonian}

We write the Hamiltonian (\ref{H-STAR}) in the form $H=H_0+H_I$,
where
\begin{equation}
 H_0 = \frac{\omega_0}{2}\sigma_3 + 2A\sigma_3 J_3
\end{equation}
is regarded as the unperturbed part and
\begin{equation}
 H_I = 2A(\sigma_+ J_- + \sigma_- J_+)
\end{equation}
as the interaction Hamiltonian. Note that the unperturbed
Hamiltonian $H_0$ is not a sum of free Hamiltonian operators for
system and environment, and that the perturbation Hamiltonian
contains only the flip-flop part of the interaction, which
describes the flip of the central spin accompanied by a back-flip
of one of the bath spins.

We transform to the interaction picture with respect to $H_0$
which leads to the interaction Hamiltonian
\begin{eqnarray*}
 H_I(t) = 2Ae^{i (\omega_0-2A)t}\sigma_+ J_-e^{4i AJ_3t} + {\mathrm{H.c.}}
\end{eqnarray*}
We will always represent our results in the interaction picture
with respect to the Hamiltonian $\frac{\omega_0}{2}\sigma_3$,
i.e.~in the rotating frame of the central spin. The results in
the interaction picture relative to $H_0$ are transformed back to
the rotating frame of the central spin through the prescription
\begin{equation}
 \rho_m(t) \longrightarrow
 e^{-2i Am\sigma_3t}\rho_m(t)e^{+2i Am\sigma_3t}.
\end{equation}
Obviously, this back-transformation leaves the populations
$P^m_{\pm}(t)$ unchanged, while the coherences pick up a phase
factor,
\begin{equation}
 \rho^m_{+-}(t) \longrightarrow e^{-4i Amt} \rho^m_{+-}(t).
\end{equation}

\subsubsection{Master equation}
For the projection defined by Eq.~(\ref{CPS1}) the NZ master
equation (\ref{NZ-RHO-I}) takes the following form in second
order,
\begin{eqnarray*}
 \lefteqn{ \frac{d}{d t} \rho_m(t) = -\sum_{m'} \int_0^t d t_1 } \\
 &\times& {\mathrm{tr}}_E
 \left\{ \Pi_{m} \left[H_I(t),\left[H_I(t_1),\rho_{m'}(t_1)\otimes
 \frac{1}{N_{m'}}\Pi_{m'} \right]\right]\right\}.
\end{eqnarray*}
The right-hand side of this equation is easily calculated to yield
\footnote{We use the convention that
$\rho_{\frac{N}{2}+1}=\rho_{-\frac{N}{2}-1}=0$.}
\begin{eqnarray} \label{MASTER-CPS1}
 \lefteqn{
 \frac{d}{d t} \rho_m(t) = 4A^2 \int_0^t d t_1 } \nonumber \\
 && \times \Big[
 \sigma_+ \rho_{m+1}(t_1)\sigma_- \Big(\frac{N}{2}+m+1\Big)
 2\cos[\Omega_+(m)(t-t_1)]
 \nonumber \\ && \;\;
 +\sigma_- \rho_{m-1}(t_1)\sigma_+ \Big(\frac{N}{2}-m+1\Big)
 2\cos[\Omega_-(m)(t-t_1)]
 \nonumber \\ && \;\;
 -\sigma_+\sigma_-\rho_{m}(t_1)\Big(\frac{N}{2}-m\Big)
 e^{i \Omega_+(m)(t-t_1)}
 \nonumber \\ && \;\;
 -\sigma_-\sigma_+\rho_{m}(t_1)\Big(\frac{N}{2}+m\Big)
 e^{i \Omega_-(m)(t-t_1)}
 \nonumber \\ && \;\;
 -\rho_{m}(t_1)\sigma_+\sigma_-\Big(\frac{N}{2}-m\Big)
 e^{-i \Omega_+(m)(t-t_1)}
 \nonumber \\ && \;\;
 -\rho_{m}(t_1)\sigma_-\sigma_+\Big(\frac{N}{2}+m\Big)
 e^{-i \Omega_-(m)(t-t_1)}\Big].
\end{eqnarray}
One easily checks that this master equation preserves the
Hermiticity and the trace of the reduced density matrix
$\rho_S(t)$ which is determined by Eq.~(\ref{RED-DENSITY}). Note
that the dynamics couple the densities $\rho_m(t)$ to $\rho_{m\pm
1}(t)$. The corresponding second-order TCL master equation
Eq.~(\ref{TCL-RHO-I}) is obtained from Eq.~(\ref{MASTER-CPS1}) by
just replacing $\rho_m(t_1)$ by $\rho_m(t)$ on the right-hand
side.

\subsubsection{Coherences}

The coherence of the central spin is given by the sum of the
coherences of the $\rho_m$,
\begin{equation}
 \rho_{+-}(t) = \sum_m \rho^m_{+-}(t).
\end{equation}
From the initial condition (\ref{INITIAL-STATE}) we find that
\begin{equation}
 \rho^m_{+-}(0) = \frac{N_m}{2^N}\rho_{+-}(0).
\end{equation}
The dynamics of the coherences $\rho^m_{+-}(t)$ is determined by
the master equation (\ref{MASTER-CPS1}) which leads to
\begin{eqnarray} \label{MASTER-COH-NZ}
 \frac{d}{d t} \rho^m_{+-}(t) &=& -\int_0^t d t_1 \left[
 B_+(m) e^{i \Omega_+(m)(t-t_1)} \right. \nonumber \\
 && \left. + B_-(m)e^{-i \Omega_-(m)(t-t_1)}
 \right] \rho^m_{+-}(t_1),
\end{eqnarray}
where
\begin{equation}
 B_{\pm}(m)=4A^2\left(\frac{N}{2}\mp m\right).
\end{equation}
Note that the dynamics does not couple the different components
$\rho^m_{+-}(t)$. The integrodifferential equation
(\ref{MASTER-COH-NZ}) can now be solved with the help of a Laplace
transformation. The corresponding TCL equation is obtained from
Eq.~(\ref{MASTER-COH-NZ}) by replacing $\rho^m_{+-}(t_1)$ by
$\rho^m_{+-}(t)$ on the right-hand side. Solving the resulting
equation yields
\begin{equation}
 \rho_{+-}(t) = \rho_{+-}(0) \sum_m \frac{N_m}{2^N} \exp\left[-4i Amt
 -\Lambda^{\mathrm{coh}}_m(t)\right],
\end{equation}
where
\begin{eqnarray}
 \Lambda^{\mathrm{coh}}_m(t)
 &=& \frac{B_+(m)}{\Omega_+^2(m)} \left[1-e^{i \Omega_+(m)t}\right]
 \nonumber \\
 && + \frac{B_-(m)}{\Omega_-^2(m)} \left[1-e^{-i \Omega_-(m)t}\right]
 \nonumber \\
 && + i \left[ \frac{B_+(m)}{\Omega_+(m)} - \frac{B_-(m)}{\Omega_-(m)}
 \right] t.
\end{eqnarray}

\begin{figure}[htb]
\begin{center}
\includegraphics[width=0.8\linewidth]{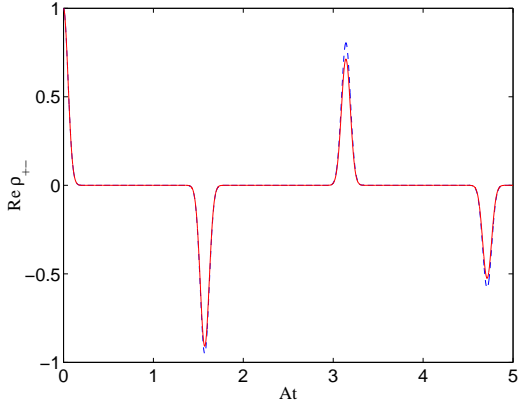}
\caption{(Color online) Coherence of the central
spin. Continuous line: Exact solution. Dashed line: Second-order
NZ approximation. Parameters: $N=101$ and $\alpha=0.1$.}
\label{fig2}
\end{center}
\end{figure}

\begin{figure}[htb]
\begin{center}
\includegraphics[width=0.8\linewidth]{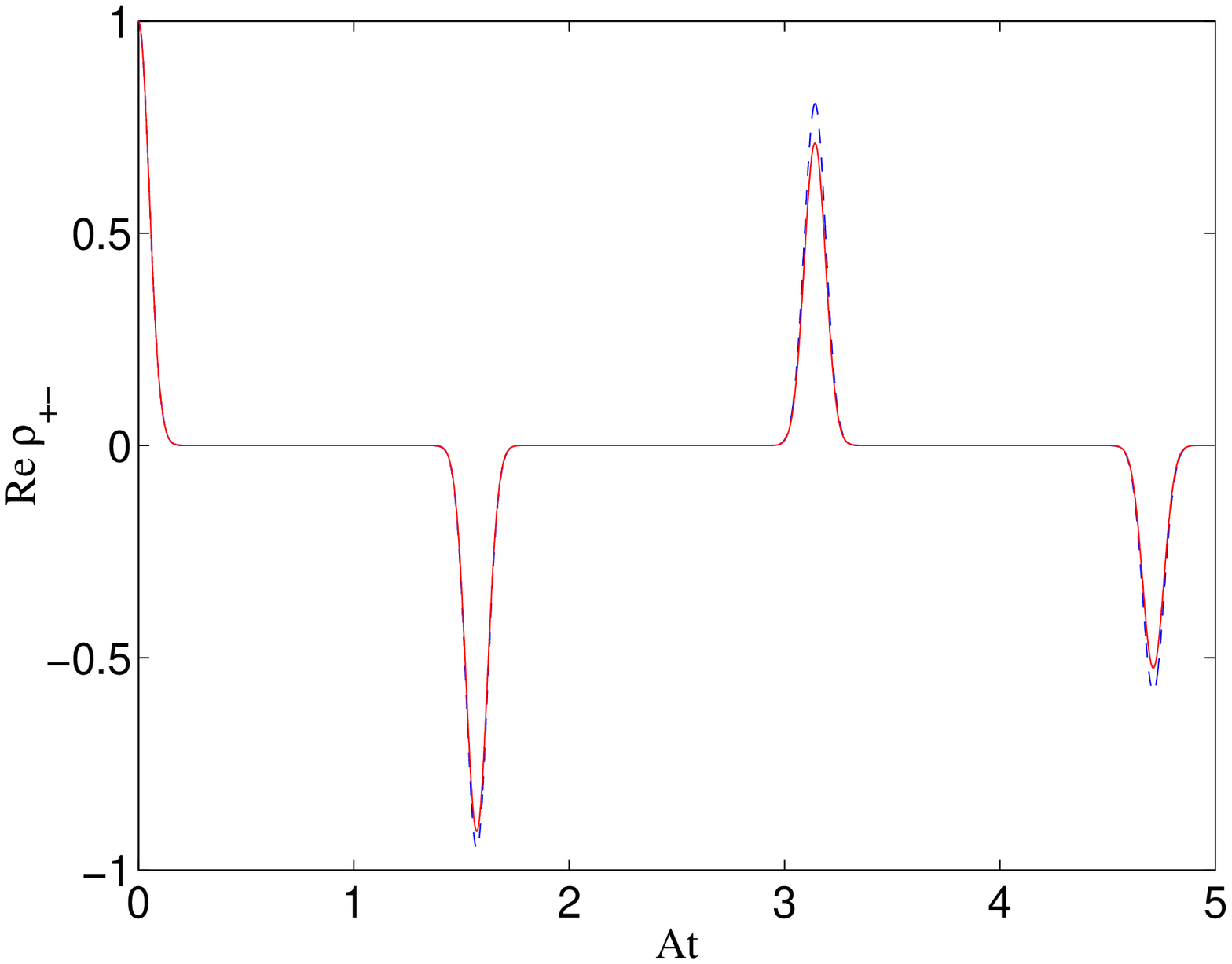}
\caption{(Color online) Coherence of the central
spin. Continuous line: Exact solution. Dashed line: Second-order
TCL approximation. Parameters: $N=101$ and $\alpha=0.1$.}
\label{fig3}
\end{center}
\end{figure}

\begin{figure}[htb]
\begin{center}
\includegraphics[width=0.8\linewidth]{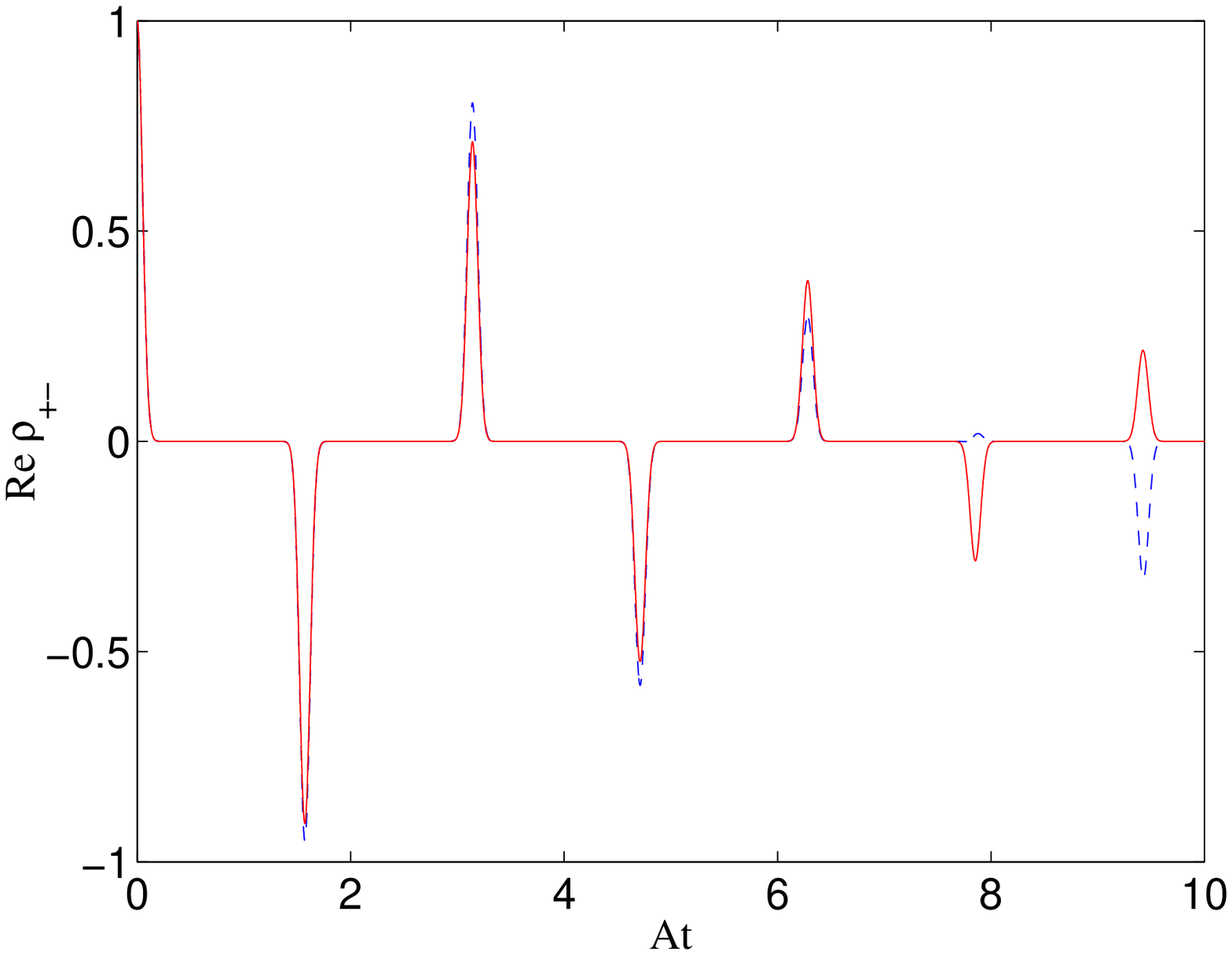}
\caption{(Color online) Coherence of the central
spin. Continuous line: Exact solution. Dashed line: Second-order
TCL approximation. Parameters: $N=101$ and $\alpha=0.1$.}
\label{fig4}
\end{center}
\end{figure}

The comparison with the exact solution shows that both methods
correspond essentially to an expansion in the dimensionless
quantity
\begin{equation}
 \alpha = \frac{2AN}{\omega_0}.
\end{equation}
We compare the results of the NZ and the TCL method with the exact
solution in Figs.~\ref{fig2} and \ref{fig3}, where the coherence
is normalized such that $\rho_{+-}(0)=1$. Both the NZ and the TCL
equation reproduce the exact solution very well in the
perturbative regime. Moreover, both are capable of describing the
position, the height and the width of the partial revivals of the
coherence, which appear because of the finite size of the bath and
the associated finite recurrence times. For large integration
times, only the positions and the width of the revivals are
reproduced correctly (see Fig.~\ref{fig4}). We note that the NZ
and the TCL method yield nearly identical results; they are hardly
distinguishable on the scale of the shown figures.

\subsubsection{Populations}

The populations are given by
\begin{equation}
 P_{\pm}(t) = \sum_m P^m_{\pm}(t).
\end{equation}
The initial condition (\ref{INITIAL-STATE}) now yields
\begin{equation}
 P^m_{+}(0) = \frac{N_m}{2^N}, \qquad P^m_{-}(0) = 0.
\end{equation}
The master equation (\ref{MASTER-CPS1}) leads to
\begin{eqnarray} \label{POP1}
 \lefteqn{
 \frac{d}{d t} P^m_+(t) = 8A^2 \int_0^t d t_1 } \nonumber \\
 && \times \left[ \left(\frac{N}{2}+m+1\right) \cos[\Omega_+(m)(t-t_1)] P^{m+1}_-(t_1)
 \right. \nonumber \\
 && \;\; - \left. \left(\frac{N}{2}-m\right) \cos[\Omega_+(m)(t-t_1)]
 P^{m}_+(t_1) \right],
\end{eqnarray}
and
\begin{eqnarray} \label{POP2}
 \lefteqn{
 \frac{d}{d t} P^m_-(t) = 8A^2 \int_0^t d t_1 } \nonumber \\
 && \times \left[ \left(\frac{N}{2}-m+1\right) \cos[\Omega_-(m)(t-t_1)] P^{m-1}_+(t_1)
 \right. \nonumber \\
 && \;\; - \left. \left(\frac{N}{2}+m\right) \cos[\Omega_-(m)(t-t_1)]
 P^{m}_-(t_1) \right].
\end{eqnarray}
Note that, in contrast to the coherences, this equation does
couple different components $P^m_{\pm}$. However, the solution can
still be constructed by noting that
\begin{equation} \label{CONS-P}
 \frac{d}{d t}\left[ P^m_+(t) + P^{m+1}_-(t) \right] = 0.
\end{equation}
With the help of this equation one shows that the total angular
momentum $J_3^{\mathrm{tot}}$ is exactly conserved under the
dynamics given by the master equation. Namely, we deduce from
Eq.~(\ref{CONS-P}) that
\begin{eqnarray}
 \lefteqn{
 \frac{d}{d t} {\mathrm{tr}}\left\{J_3^{\mathrm{tot}}{\mathcal{P}}\rho(t)\right\}
 } \nonumber \\
 && = \frac{d}{d t} \sum_m \left(\frac{1}{2}+m\right)
 \left[P_+^m(t)+P_-^{m+1}(t) \right] = 0.
\end{eqnarray}

\begin{figure}[htb]
\begin{center}
\includegraphics[width=0.8\linewidth]{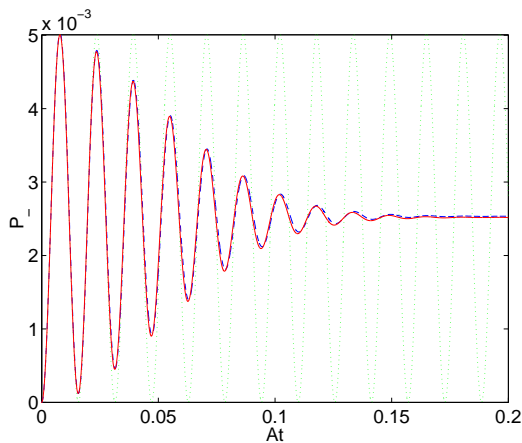}
\caption{(Color online) Population of the central spin according
to the exact solution (continuous line), and to the second-order
TCL approximations using the correlated projection (broken line)
and the standard product-state projection (dotted line).
Parameters: $N=101$ and $\alpha=0.5$.} \label{fig5}
\end{center}
\end{figure}

\begin{figure}[htb]
\begin{center}
\includegraphics[width=0.9\linewidth]{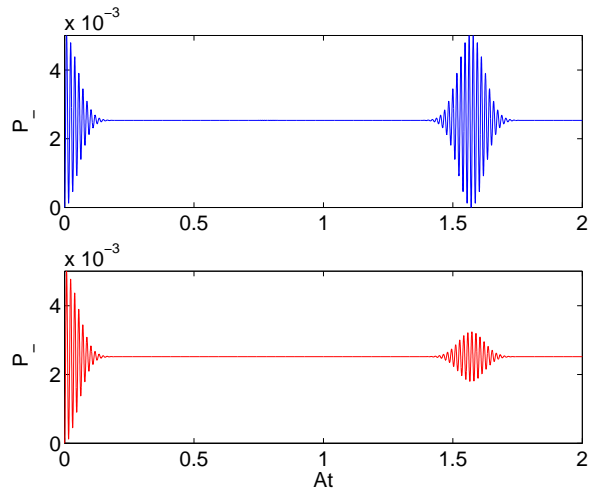}
\caption{(Color online) Population of the central
spin according to the second-order TCL approximation (top) and to
the exact solution (bottom). Parameters: $N=101$ and
$\alpha=0.5$.} \label{fig6}
\end{center}
\end{figure}

Equation (\ref{CONS-P}) is used to eliminate $P^{m+1}_-$ from the
right-hand side of Eq.~(\ref{POP1}) to get a closed equation for
$P_+^m$ which can again be solved with the help of a Laplace
transformation. The corresponding TCL equations are solved in a
similar manner yielding
\begin{equation}
 P_+(t) = \sum_m \frac{N_m}{2^N} \left[
 \frac{\frac{N}{2}+m+1}{N+1} + \frac{\frac{N}{2}-m}{N+1}
 e^{-\Lambda^{\mathrm{pop}}_m(t)}
 \right],
\end{equation}
where
\begin{equation}
 \Lambda^{\mathrm{pop}}_m(t) = \frac{8A^2(N+1)}{\Omega_+^2(m)}
 \left( 1-\cos[\Omega_+(m)t] \right).
\end{equation}
As demonstrated in Fig.~\ref{fig5} this result gives an excellent
approximation of the exact dynamics in the perturbative regime.
For longer integration times, the position and the width of the
revivals are reproduced very well, while their height is not (see
Fig.~\ref{fig6}). We find again that the NZ result is hardly
distinguishable from the corresponding TCL result, and hence
refrain from showing it in a separate figure.

The standard procedure that uses the product-state projection
operator (\ref{STANDARD-PROJECTION}) with $\rho_0=2^{-N}I_E$
yields the upper state population
\begin{equation}
 P_+(t) = \frac{1}{2} \left[
 1 + \exp\left(-\frac{8A^2N}{\omega_0^2}(1-\cos\omega_0t)\right)
 \right],
\end{equation}
which is also plotted in Fig.~\ref{fig5}. We see that the standard
projection operator technique leads to a rather bad approximation
which is neither able to represent correctly the damping nor the
revivals of the populations.

\subsection{Complete projection onto the angular momentum manifolds}
For the present model we can construct another correlated
projection superoperator which takes into account both conserved
quantities, namely $J_3^{\mathrm{tot}}$ and $\bm{J}^2$. This
projection is given by
\begin{equation} \label{PROJECTION-JM}
 {\mathcal{P}}\rho = \sum_{jm} {\mathrm{tr}}_E \{ \Pi_{jm} \rho\}
 \otimes \frac{1}{N_j} \Pi_{jm}
 \equiv \sum_{jm} \rho_{jm} \otimes \frac{1}{N_j} \Pi_{jm},
\end{equation}
where $N_j$ has already been defined in Eq.~(\ref{Def-N-J}). The
$\Pi_{jm}$ denote the ordinary projection operators that project
onto the common eigenspaces of $J_3$ and $\bm{J}^2$ with
corresponding eigenvalues $m$ and $j(j+1)$. The map
(\ref{PROJECTION-JM}) obviously leaves invariant
$J_3^{\mathrm{tot}}$ and $\bm{J}^2$ and fulfills again all
conditions formulated in Sec.~\ref{SEC:CORR}.

The second-order NZ master equation corresponding to the
projection (\ref{PROJECTION-JM}) reads
\begin{eqnarray} \label{MASTER-CPS2}
 \lefteqn{
 \frac{d}{d t} \rho_{jm}(t) = 4A^2 \int_0^t d t_1 } \nonumber \\
 && \times \Big[
 \sigma_+ \rho_{j,m+1}(t_1)\sigma_- b(j,m) 2\cos[\Omega_+(m)(t-t_1)]
 \nonumber \\ && \;\;
 +\sigma_- \rho_{j,m-1}(t_1)\sigma_+ b(j,-m) 2 \cos[\Omega_-(m)(t-t_1)]
 \nonumber \\ && \;\;
 -\sigma_+\sigma_-\rho_{jm}(t_1) b(j,m) e^{i \Omega_+(m)(t-t_1)}
 \nonumber \\ && \;\;
 -\sigma_-\sigma_+\rho_{jm}(t_1) b(j,-m) e^{i \Omega_-(m)(t-t_1)}
 \nonumber \\ && \;\;
 -\rho_{jm}(t_1)\sigma_+\sigma_- b(j,m) e^{-i \Omega_+(m)(t-t_1)}
 \nonumber \\ && \;\;
 -\rho_{jm}(t_1)\sigma_-\sigma_+ b(j,-m) e^{-i \Omega_-(m)(t-t_1)}\Big],
\end{eqnarray}
where $b(j,m)=j(j+1)-m(m+1)$. Of course, the TCL version is
obtained again by replacing $\rho_{jm}(t_1)$ with $\rho_{jm}(t)$
under the time integral.

\begin{figure}[htb]
\begin{center}
\includegraphics[width=0.9\linewidth]{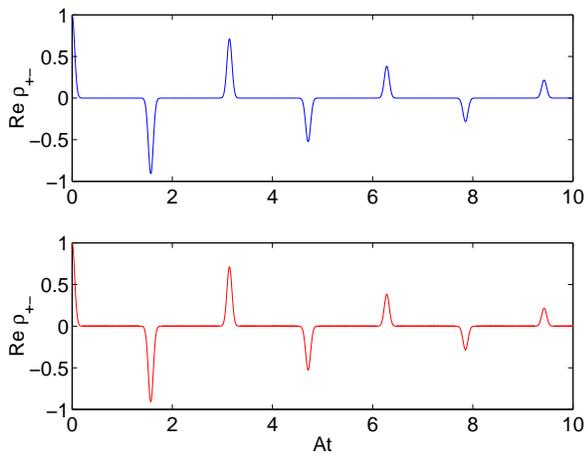}
\caption{(Color online) Coherence of the central spin. Top:
Second-order TCL approximation for the projection
(\ref{PROJECTION-JM}). Bottom: Exact solution. Parameters: $N=101$
and $\alpha=0.1$.} \label{fig7}
\end{center}
\end{figure}

The procedure for the determination of the coherences and the
populations of the central spin is analogous to the one of the
previous case. Again, we find that the NZ and the TCL approach
yield similar results for the coherences. An example is shown in
Fig.~\ref{fig7} for the TCL approximation only. We see that the
second order with the projection (\ref{PROJECTION-JM}) leads to an
excellent approximation for the coherences even for very long
integration times.

A further remarkable feature of the projection
(\ref{PROJECTION-JM}) is that the corresponding NZ equation of
second order (\ref{MASTER-CPS2}) reproduces the \textit{exact}
dynamics of the populations. This can be demonstrated by solving
the equations for the populations obtained from the master
equation (\ref{MASTER-CPS2}). Alternatively, we can prove this
statement directly with the help of the following argument. We
first note that by use of Eq.~(\ref{PLP-ASSUM}) the von Neumann
equation (\ref{NEUMANN-EQ}) leads to the exact equation
\begin{equation}
 \frac{d}{d t} {\mathcal{P}}\rho(t) = \int_0^t d t_1
 {\mathcal{P}}{\mathcal{L}}(t){\mathcal{L}}(t_1)\rho(t_1),
\end{equation}
from which we obtain an exact equation for the populations of the
central spin:
\begin{equation} \label{POP-EXACT}
 \frac{d}{d t}\langle\pm|{\mathcal{P}}\rho(t)|\pm\rangle = \int_0^t d t_1
 \langle\pm|{\mathcal{P}}{\mathcal{L}}(t){\mathcal{L}}(t_1)\rho(t_1)|\pm\rangle.
\end{equation}
The crucial point is that for arbitrary $\rho(t_1)$ the following
relation holds,
\begin{equation} \label{POP-REL}
 \langle\pm|{\mathcal{P}}{\mathcal{L}}(t){\mathcal{L}}(t_1)\rho(t_1)|\pm\rangle =
 \langle\pm|{\mathcal{P}}{\mathcal{L}}(t){\mathcal{L}}(t_1){\mathcal{P}}\rho(t_1)|\pm\rangle.
\end{equation}
Hence, when calculating the diagonal elements of the density
matrix we may insert at any time a factor of ${\mathcal{P}}$ in
front of the density matrix $\rho(t_1)$. This fact is obviously
related to the invariance of the subspaces spanned by the states
$|+\rangle\otimes|j,m\rangle$ and $|-\rangle\otimes|j,m+1\rangle$,
a fact that has already been used to solve the Schr\"odinger
equation. For a formal proof one uses the expression
(\ref{PROJECTION-JM}) to demonstrate that Eq.~(\ref{POP-REL}) is a
direct consequence of the relation
\begin{equation}
 {\mathcal{L}}^{\dagger}(t_1){\mathcal{L}}^{\dagger}(t)
 \left( |\pm\rangle\langle\pm| \otimes \Pi_{jm} \right) =
 \sum_{m'} A_{jm'} \otimes \Pi_{jm'}
\end{equation}
with certain system operators $A_{jm'}$.

Using now the relation (\ref{POP-REL}) in Eq.~(\ref{POP-EXACT}) we
find
\begin{equation}
 \frac{d}{d t}\langle\pm|{\mathcal{P}}\rho(t)|\pm\rangle = \int_0^t d t_1
 \langle\pm|{\mathcal{P}}{\mathcal{L}}(t){\mathcal{L}}(t_1){\mathcal{P}}\rho(t_1)|\pm\rangle.
\end{equation}
This equation is exact and coincides with the equation for the
population that is obtained from the second-order NZ equation
(\ref{NZ2}). We conclude that the NZ equation of second order
indeed yields the exact dynamics of the populations.

\section{Conclusions}\label{CONCLU}

We have investigated the performance of the correlated projection
operator method in the case of a spin-bath model where a central
spin couples to a spin bath via full Heisenberg interaction. The
coupling constants have been chosen to be uniform such that one
can construct an exact solution of the von Neumann equation. We
have considered two different projection superoperators, one of
which projects onto the eigenspaces of the $3$-component $J_3$ of
the total bath angular momentum. The other projection operator
projects onto the simultaneous eigenspaces of $J_3$ and the square
$\bm{J}^2$ of the total bath angular momentum. The choice of these
projections is motivated by the fact that both
$J_3^{\mathrm{tot}}$ and $\bm{J}^2$ are conserved under the
dynamics given by the full system Hamiltonian. The projection
superoperators studied here thus fully exploit the known
symmetries of the model, granting the invariance of the
expectation values of the conserved quantities. For both
projectors we have solved the Nakajima-Zwanzig and the
time-convolutionless master equation to second order in the
interaction and have compared the results to the exact solution.

Both the NZ and the TCL master equation lead to a relaxation and
decoherence behavior that approximates the exact solution very
well. Remarkably, also the position and the width of the partial
revivals of the coherence and the populations could be reproduced.
For all tested parameter sets, the TCL and NZ solutions were
hardly distinguishable. In these cases the TCL approach is to be
favored because of its much simpler structure.

A remarkable property of the projection superoperator that
projects onto the simultaneous eigenspaces of $\bm{J}^2$ and $J_3$
is that the corresponding second-order NZ equation results in the
exact dynamics for the populations. While the standard product
state projection gives only a poor approximation, the above
correlated projection reproduces the exact solution already in
lowest order. This illustrates an important feature of the
correlated projection operator method. Namely, different choices
for the projection ${\mathcal{P}}$ yield dynamic equations of
different structure for different sets of relevant variables.
Therefore, the resulting expressions for the quantities of
interest, e.g.~the populations, may differ in all orders of the
coupling. Hence, changing the projection operator corresponds in
general to a non-perturbative reorganization of the expansion for
the desired quantities.

Finally, we emphasize that all calculations in this paper
involving projection operator techniques can be carried out in a
similar way for more realistic assumptions regarding the coupling
constants. The correlated projection operator technique may thus
be applicable to many physically relevant problems featuring
non-Markovian dynamics, such as the hyperfine interaction of an
electron confined to a quantum dot, the spin dynamics under a
spin-echo pulse sequence, or the spin dynamics in a molecular
magnet.

\begin{acknowledgments}
J.F. acknowledges support from the Swiss NF and the NCCR
Nanoscience, and thanks W. A. Coish for helpful comments on the
manuscript.
\end{acknowledgments}

\end{document}